\documentstyle[prl,aps,preprint]{revtex}
\begin{document}
\draft
\title{Separability of mixed states: \\
necessary and sufficient conditions}

\author{Micha\l{}   Horodecki}

\address{Departament of Mathematics and Physics\\
 University of Gda\'nsk, 80--952 Gda\'nsk, Poland}

\author{Pawe\l{} Horodecki}

\address{Faculty of Applied Physics and Mathematics\\
Technical University of Gda\'nsk, 80--952 Gda\'nsk, Poland}

\author{Ryszard Horodecki\footnote{e-mail: fizrh@univ.gda.pl} }

\address{Institute of Theoretical Physics and Astrophysics\\
University of Gda\'nsk, 80--952 Gda\'nsk, Poland}

\maketitle

\begin{abstract}
We provide  necessary and sufficient conditions for separability of mixed
states.
As a result we obtain a simple criterion of separability for
$2\times2$  and $2\times3$ systems. Here, the positivity of the
partial transposition of a state is necessary and sufficient for its
separability. However, it is not the case in general. Some examples of
mixtures which demonstrate the utility of the criterion are considered.
\end{abstract}

\newtheorem{theorem}{Theorem}
\newtheorem{lemma}{Lemma}
Quantum inseparability, first recognized  by Einstein, Podolsky and Rosen
\cite{EPR} and Schr\"odinger \cite{Schr}, is one of the most astonishing
features of quantum formalism. After over sixty years it is still a
fascinating object from both theoretical and experimental points of view.
Recently, together with a dynamical development of
experimental methods, a number of possible practical applications of
the quantum inseparable states has been proposed including quantum
computation \cite{Deutsch} and quantum teleportation \cite{Bennett}.
The above ideas are based on the fact
that the quantum inseparability implies, in particular, the  existence of the
pure entangled states which
produce nonclassical phenomena. However, in
laboratory one deals with mixed states rather than pure ones.  This is  due to
the uncontrolled interaction with the enviroment. Then it is very important
to know which mixed states can produce quantum effects.
The problem is much more complicated than in the pure states case. It may be
due to the fact that mixed states apparently possess the ability to behave
classically in some respects but quantum mechanically in others 
\cite{Popescu94}.

In accordance with the  so-called generalized inseparability principle
\cite{inf} we will call  a mixed state of a compound quantum system  inseparable
if it cannot be written as a convex
combination of product states. The problem of the inseparability of mixed 
states was first raised by Werner \cite{Werner}, who constructed  a family of
inseparable
states which admit the local hidden variable model. It has been pointed
out \cite{Popescu95} that, nevertheless, some of them are nonlocal
and this ``hidden'' nonlocality can be
revealed by  subjecting them  to more complicated experiments than single von
Neumann measurements considered by Werner (see also Ref. \cite{Gisin} in this
context).
This shows that it is hard to
classify the mixed states as purely classical or quantum ones.

Recently the separable states have been investigated within the
information-theoretic approach \cite{inf,red,renyi,prep}. It has been shown
that they satisfy a series
of the so-called quantum $\alpha$-entropy inequalities (for $\alpha=1,2$
\cite{red,renyi} and $\alpha=\infty$ \cite{prep}). Moreover, the separable
two spin-$1\over2$ states with maximal entropies of subsystems have been
completely characterized in terms of the $\alpha$-entropy inequalities 
\cite{inf}.
It is remarkable that there exist inseparable states which do {\it not} reveal
nonclassical features under the entropic criterion \cite{renyi}.

Then the fundamental problem of
an ``operational'' characterization of the separable states arises.
So far only some necessary conditions of separability have been found
\cite{inf,Werner,red,renyi,mybell}.
An important step is due to Peres
\cite{Peres}, who has provided a very strong condition. Namely, he noticed that
the separable states remain positive if subjected to partial transposition.
Then he conjectured that this is  also sufficient condition.

In this Letter we present  two necessary and sufficient  conditions for
separability of mixed
states. It provides a complete, operational characterization of
separable states for $2\times2$ and $2\times3$ systems. It appears that
Peres' conjecture is valid for those cases. However,
as we show in the Appendix, the conjecture is not valid in general.
We also illustrate our results by means of some examples and discuss 
them in the context of the $\alpha$-entropy inequalities.

To make our considerations more clear, we start from the following
notation and definitions.
 We will deal with the states  on the
finite dimensional Hilbert space ${\cal H}={\cal H}_1\otimes {\cal H}_2$.
An operator $\varrho$ acting on $\cal H$ is a state if $\text{Tr}\varrho=1$
and if it is a positive operator i.e.
\begin{equation}
\text{Tr} \varrho P\geq0
\end{equation}
for any projectors $P$. A state is called separable \footnote{The
presented definition of separable states is due to Werner \cite{Werner} who
called them classically correlated states.}
if it can be approximated
in the trace norm by the states of the form
\begin{equation}
\varrho=\sum_{i=1}^kp_i\varrho_i\otimes\tilde \varrho_i
\end{equation}
where $\varrho_i$ and $\tilde\varrho_i$ are states on ${\cal H}_1$ and
${\cal H}_2$ respectively.
By ${\cal A}_1$ and ${\cal A}_2$ we will denote the set of the operators acting
on ${\cal H}_1$ and ${\cal H}_2$ respectively. Recall that ${\cal A}_i$
constitute a Hilbert space (so-called Hilbert-Schmidt space)
with scalar product $\langle A,B\rangle=\text{Tr}B^\dagger A$.
The space of the linear maps from ${\cal A}_1$ to ${\cal A}_2$ is denoted by
${\cal L}({\cal A}_1,{\cal A}_2)$. We say that a map
$\Lambda\in{\cal L}({\cal A}_1,{\cal A}_2)$
 is positive if it maps positive operators
in ${\cal A}_1$ into the set of positive operators i.e.  if $A\geq0$ implies
$\Lambda(A)\geq0$. Finally we need the definition of completely positive map.
One says \cite{Lind} that a map $\Lambda\in{\cal L}({\cal A}_1,{\cal A}_2)$ is
completely positive if the induced map
\begin{equation}
\Lambda_n=\Lambda\otimes I:{\cal A}_1\otimes {\cal M}_n \rightarrow
{\cal A}_2\otimes {\cal M}_n
\end{equation}
is positive for all $n$. Here ${\cal M}_n$ stand for the set of the complex
matrices $n\times n$ and $I$ is the identity map\footnote{Of course a
completely positive map is also a positive one.}.

Thus the tensor product of a completely positive map and the identity
maps positive
operators into positive ones. It is remarkable that there are positive maps
that do not possess this property. This fact is crucial for the problem we
discuss here. Indeed, trivially, the product states are mapped into positive
operators by the tensor product of a positive map and identity:
$(\Lambda\otimes I)\varrho\otimes\tilde\varrho=
(\Lambda\varrho)\otimes\tilde\varrho\geq0$. Of course,  the same holds
for the separable
states. Then our main idea is  that this property of the separable
states is essential i.e., roughly speaking, if a state $\varrho$ is
inseparable, then there exists a
positive map $\Lambda$ such that $\Lambda\otimes I\varrho$ is {\it not}
positive.
This means that  we can recognize the inseparable states by means of the 
positive maps.
Now  the point is that not all the positive maps can help us to determine
whether a given state is inseparable. In fact,  the completely positive
maps do not ``feel'' the inseparability.
Thus the problem of characterization of  the set of the separable states
reduces to
the following: one should extract from thet set of all positive maps
some essential ones.
As we will see further, it is possible in some cases. Namely it appears that
for the $2\times2$ and $2\times3$ systems the transposition is the {\it only}
such map.

We will start from  the following
\begin{lemma}
For any inseparable state $\tilde\varrho\in {\cal A}_1\otimes{\cal A}_2$
there exists Hermitian operator $\tilde{A}$ such that
\begin{equation}
Tr(\tilde{A}\tilde{\varrho})<0 \ \   \mbox{and} \ \  Tr(\tilde{A}{\sigma})\geq 0
\end{equation}
for any separable  $\sigma$.
\label{lem}
\end{lemma}

{\it Proof.-}
From the  definition  of the set of separable states it follows that it is
both convex and closed set in ${\cal A}_1\otimes{\cal A}_2$.
Thus we can apply a theorem (conclusion from the Hahn-Banach theorem)
\cite{hb} which, for our purposes, can be formulated as follows.
If  $W_1$, $W_2$ are convex closed sets in  a real Banach space
and one of them is compact, then there exists
a continous functional $f$ and $\alpha\in R$ such that
for all pairs $w_1 \in W_1$, $w_2 \in W_2$ we have
\begin{equation}
f(w_1)<\alpha \leq f(w_2)
\end{equation}
This theorem says, in particular, that a closed convex set in
the Banach space is
completely described by the inequalities involving continous functionals.

Noting that one-element set is compact we obtain that there exists a real
functional $g$ on the real space $\tilde {\cal A}$ generated by
Hermitian operators from ${\cal A}_1\otimes{\cal A}_2$ such that
\begin{equation}
g(\tilde \varrho)<\beta \leq g(\sigma),
\end{equation}
for all separable $\sigma$.
It is a well known fact that any contionous functional g
on a Hilbert space can be represented by a vector from this space. As
$\tilde {\cal A}$ is a (real)  Hilbert space we obtain that
the functional $g$ can be represented as
\begin{equation}
g(\varrho)=Tr(\varrho A),
\end{equation}
where $A=A^\dagger$.
Now, if $I$ stands for identity then
for any states $\varrho$, $\sigma$ one has obviously
$Tr(\beta I \varrho)= Tr(\beta I \sigma)=\beta$.
Thus it sufficies only to take
\begin{equation}
\tilde{A}=A-\beta I
\end{equation}
to complete the proof of the lemma.
The lemma allows us to prove the following
\begin{theorem}
A state $\varrho\in{\cal A}_1\otimes{\cal A}_2 $ is separable iff
\begin{equation}
Tr(A\varrho) \geq 0
\label{eqfunk}
\end{equation}
for any Hermitian operator $A$ satisfying $Tr(A P\otimes Q) \geq 0$,
where P and Q are projections acting on ${\cal H}_1$ and ${\cal H}_2$
respectively.
\label{funk}
\end{theorem}

{\it Proof .-} If $\varrho$ is separable then obviously
it satisfies the condition (\ref{eqfunk}). To prove the converse statement,
suppose that  $\varrho$ satisfies the condition  (\ref{eqfunk})
and is inseparable.
Then, due to inseparability,
it would be possible (on the strength of the lemma)
to find  some Hermitian operator $A$
for which  $Tr(A \varrho) < 0$ although
$Tr(A \sigma) \geq 0$ for any separable state $\sigma$ or, equivalently, for
any product projector $P\otimes Q$ which is a contradiction.

{\it Remark 1.-} 
Note that the if an operator $A$ satisfies the condition 
$\text{Tr}AP\otimes Q\geq0$ for any projectors $P$ and $Q$
then it is Hermitian. 

{\it Remark 2.-}
As the conclusion from the Hahn-Banach theorem
is valid for {\it any} Banach space our
theorem can be generalized  for infinitely dimensional Hilbert spaces.
Namely, it can be seen that
the condition is $\text{Tr} A\varrho\geq0$  for any bounded $A$ such that
$\text{Tr}AP\otimes Q\geq0$ for any projectors $P$ and $Q$
\footnote{It follows from the fact
that the set of the continuous functionals of the Banach space of trace class
operators is isomorphic  to the set of bounded operators.}.

A good example of a nontrivial operator which satisfies the condition
(\ref{eqfunk}) is the one used by Werner \cite{Werner}, defined on
$C^d\otimes C^d$ by $V\phi\otimes\tilde\phi=\tilde\phi\otimes\phi$.
One can check that $\text{Tr}VP\otimes Q=\text{Tr}PQ\geq0$. This operator,
together with its $U_1\otimes U_2$ transformations allowed to characterize
the set of separable states within the class of two spin-$1\over 2$ mixtures
with maximal entropies of the subsystems \cite{inf}.

Now the main task is to translate the above theorem into the language of
positive maps. For this purpose we will use the  isomorphism between the
positive maps and operators which are positive on the product projectors. Namely
an arbitrarily established orthonormal basis in ${\cal A}_1$ defines an
isomorphic map $\cal S$
from the set of linear maps $\Lambda:{\cal A}_1\rightarrow{\cal A}_2$
into operators acting on ${\cal H}_1\otimes {\cal H}_2$:

\begin{equation}
{\cal L}({\cal A}_1,{\cal A}_2)\ni \Lambda \rightarrow
{\cal S}(\Lambda)=\sum_iE_i^\dagger\otimes\Lambda(E_i)\in
{\cal A}_1\otimes {\cal A}_2
\end{equation}
According to \cite{Jamiol} a transformation
$\Lambda\in {\cal L}({\cal A}_1,{\cal A}_2)$ is positive iff
${\cal S}(\Lambda)$ is Hermitian and
$\text{Tr}\bigl({\cal S}(\Lambda)P\otimes Q\bigr)\geq0$ for any projectors
$P\in{\cal A}_1$ and $Q\in{\cal A}_2$.

Now, let us take as a basis in
${\cal A}_1$ the set of the operators $\{P_{ij}\}_{i,j=1}^{\dim {\cal H}_1}$
given by $P_{ij}e_l=\delta_{jl} e_i$ for any established basis $\{e_l\}$ in
${\cal H}_1$.
Then the condition (\ref{eqfunk}) is equivalent to the following one
\begin{equation}
\text{Tr} \left\{\left[(I\otimes \Lambda)\smash{\sum_{\smash{ij}}}P_{ji}\otimes
P_{ij}\right]\varrho\right\}\geq0
\end{equation}
or
\begin{equation}
\text{Tr} \left\{\left[(I\otimes \Lambda T)\smash{\sum_{\smash{ij}}}P_{ji}
\otimes TP_{ij}\right]\varrho\right\}\geq0
\label{cond2}
\end{equation}
where $T:{\cal A}_1\rightarrow {\cal A}_1$ is given by $TP_{ij}=P_{ji}$ i.e.
it is transposition of the operator written in a basis $\{e_i\}$.
Of course $T$ is a positive map and $T^2=I$. Then any positive map
$\tilde\Lambda:{\cal A}_1\rightarrow {\cal A}_2$ is of the
form $T\Lambda$ where $\Lambda$ is a positive map.
Putting $P_0={1\over d}\sum_{ij}P_{ji}\otimes P_{ji}$ where $d=\text
{dim}{\cal H}_1$, and using the scalar product in Hilbert space
${\cal A}_1\otimes {\cal A}_2$ the condition (\ref{cond2}) can be rewritten
in the form
\begin{equation}
\langle \varrho,(I\otimes\Lambda P_0)^\dagger \rangle\geq0.
\end{equation}
However, positive maps preserve Hermiticity hence the tensor product
$I\otimes\Lambda$ also does. Then, as $P_0$ is Hermitian we obtain
\begin{equation}
\langle \varrho,I\otimes\Lambda P_0 \rangle \geq0.
\end{equation}
This is equivalent, by passing to the adjoint maps, to the condition
\begin{equation}
\langle I\otimes\Lambda\varrho,P_0\rangle\equiv
\text{Tr}[P_0 (I\otimes \Lambda\varrho)] \geq0.
\label{war}
\end{equation}
for any positive maps $\Lambda:{\cal A}_2\rightarrow {\cal A}_1$.
Now, if a state is separable, then obviously the operator
$I\otimes \Lambda\varrho$ is positive
for any positive $\Lambda$. Conversly, if $I\otimes \Lambda\varrho$ is positive
for any $\Lambda$, then as $P_0$ is a (one dimensional) projector,
the condition (\ref{war}) is satisfied hence the
state is separable.  In this way we have obtained the main result of this
paper

\begin{theorem}
Let $\varrho$ act on Hilbert space ${\cal H}_1 \otimes {\cal H}_2$. Then
$\varrho$ is separable iff for any positive map
$\Lambda:{\cal A}_2\rightarrow {\cal A}_1$ the operator
$I\otimes\Lambda\varrho$ is positive.
\label{dod}
\end{theorem}

On the basis of the above theorem one can  characterize the set of the
separable states for $2\times2$ and $2\times3$ systems.
Namely	we have

\begin{theorem}
A state $\varrho$ acting on $C^2\otimes C^2$ or $C^2\otimes C^3$ is separable
iff its partial transposition is a positive operator.
\end{theorem}
Here the partial transposition $\varrho^{T_2}$ is given by
\begin{equation}
\varrho^{T_2}=I\otimes T\varrho
\label{transp}
\end{equation}

{\it Proof.-}
If $\varrho$ is separable then of course $\varrho^{T_2}$ is positive\footnote{%
This was first proved by Peres \cite{Peres}. In our approach it is
a consequence of the fact that $T$ is a positive map.}. To prove the
converse statement we will use Str\o{}mer and Woronowicz results
\cite{Stromer,Woronowicz}. Namely
the authors showed that any positive map $\Lambda:{\cal
A}_1\rightarrow{\cal A}_2$ with ${\cal H}_1={\cal H}_2=C^2$ \cite{Stromer}
or ${\cal H}_1=C^3$, ${\cal H}_2=C^2$ (equivalently
${\cal H}_1=C^2$, ${\cal H}_2=C^3$)
\cite{Woronowicz} is of the form
\begin{equation}
\Lambda=\Lambda^{CP}_1+\Lambda^{CP}_2T
\label{cp}
\end{equation}
where $\Lambda_i^{CP}$ are completely positive maps.
Now due to the complete positivity of $\Lambda^{CP}_i$ the map
$\Lambda_i=I\otimes\Lambda^{CP}_i$ is a positive one.
If $\varrho^{T_2}$ is positive then $\Lambda_1\varrho+\Lambda_2\varrho^{T_2}$
also
does hence from the Theorem \ref{dod} it follows that $\varrho$ is separable.

{\it Remark .-} It is easy to see that in the Theorem 2 one can
put $\tilde{\Lambda} \otimes \Lambda$ or $\Lambda \otimes I$ instead of
$I \otimes \Lambda$
(involving any positive $\tilde{\Lambda}:A_1 \rightarrow A_2$,
$\Lambda: A_2 \rightarrow A_1$).
Consequently one can use the condition of positivity
of partial transposition with respect to the first space \cite{Peres}.
The condition involving the second space is used here
only for convenience.

The above theorem is an important result, as it allows us to determine
unambigiously whether a given quantum state of $2\times2$ ($2\times 3$)
system can be written as mixture of product states or not.
It follows that for the considered cases Peres' conjecture {\it is
valid}. Hence the necessary and sufficient condition is here surprisingly
simple.
However we will see that it is not true in general.
Namely   the positive maps \cite{Woronowicz}
are characterized by the formula (\ref{cp}) only for the considered cases.
In the Appendix we present the  outline of the proof that the condition
$\varrho^{T_2}\geq0$ is, in general, only a necessary one.

{\it Examples .-} We will use the following notation for matrix elements of any
state of $M \times N$ system
( i.e. of any density matrix acting on $C^M \otimes C^N$
space) (c.f. \cite{Peres}):
\begin{eqnarray}
\varrho_{m \mu, n \nu}=\langle e_m \otimes f_{\mu} | \varrho|
e_n \otimes f_{\nu} \rangle,
\end{eqnarray}
where $\{e_m\}$ ($\{f_{\mu}\}$) denotes the arbitrary
orhonormal basis in Hilbert space describing
first (second) system.
Hence the partial transposition of $\varrho$ is defined as:
\begin{eqnarray}
\varrho^{T_2}_{m \mu, n \nu}\equiv \varrho_{m \nu, n \mu}.
\end{eqnarray}
On the other hand the $M \times N$ state
can be written as
\begin{eqnarray}
\varrho=\left[ \begin{array}{ccc}
         A_{11} \ ... \  A_{1M} \\
         ... \ ... \ ... \\
         A_{M1} \ ... \ A_{MM} \\
       \end{array}
      \right ],
\end{eqnarray}
with $N\times N$ matrices
$A_{m n}$  acting on the second ($C^N$) space.
They are defined by their matrix elements as
$\{ A_{m n} \}_{\mu \nu} \equiv  \varrho_{m \nu, n \mu}$.
Then the partial transposition will be realised simply by transposition
of all of these matrices, namely
\begin{eqnarray}
\varrho=\left[ \begin{array}{ccc}
         A_{11}^T \ ... \  A_{1M}^T \\
         ... \ ... \ ... \\
         A_{M1}^T \ ... \ A_{MM}^T \\
       \end{array}
      \right ].
\end{eqnarray}
Consider now the examples of the two spin-${1 \over 2}$ states.
The following states has been introduced
in the context of inseparability and Bell inequalities \cite{tata}
\begin{equation}
\varrho=p|\psi_1\rangle\langle\psi_1| +
(1-p)|\psi_2\rangle\langle\psi_2|
\label{st}
\end{equation}
where
$|\psi_1\rangle =ae_1\otimes e_1+be_2\otimes e_2  $,
$|\psi_2\rangle =ae_1\otimes e_2+be_2\otimes e_1$
with $a,b>0$, $\{e_i\}$ being standard basis in $C^2$.
In this case we have
\begin{eqnarray}
\varrho=\left[ \begin{array}{cccc}
          p a^2 & 0          & 0         & p ab    \\
           0    & (1-p) a^2  & (1-p) ab  & 0       \\
           0    & (1-p) ab   & (1-p) b^2 & 0       \\
          p ab  &  0         &  0         & p  b^2
       \end{array}
      \right ], \ \ \ \nonumber\\
\varrho^{T_2}=\left[ \begin{array}{cccc}
          p a^2 & 0          & 0         & (1-p) ab    \\
           0    & (1-p) a^2  & p ab  & 0       \\
           0    & p ab   & (1-p) b^2 & 0       \\
         (1-p) ab  & 0          & 0         & p  b^2
       \end{array}
      \right ].
\end{eqnarray}
We will investigate the condition of positivity
of the partial transposition $\varrho^T_2$.
For this purpose one needs to check the following determinants
\begin{eqnarray}
W_1=\varrho^{T_2}_{11,11}\varrho^{T_2}_{22,22}-
\varrho^{T_2}_{11,22}\varrho^{T_2}_{22,11} \nonumber \\
W_2=\varrho^{T_2}_{12,12}\varrho^{T_2}_{21,21}-
\varrho^{T_2}_{12,21}\varrho^{T_2}_{21,12}
\label{wyzn}
\end{eqnarray}
We obtain $W_1= a^2 b^2(2p-1)$ and $W_2=-W_1$. Thus for
$ab\neq 0$, $p\neq {1 \over 2}$ one of the determinants is negative
hence by virtue of the necessity of the condition (\cite{Peres},
Theorem 3) we conclude that $\varrho$ is inseparable for such parameters.
For a=b both  $|\psi_1\rangle$ and  $|\psi_2\rangle$ become
product states and then $\varrho$ is separable.
Finally, for $p={ 1 \over 2}$ it apears that
$\varrho^{T_2}=\varrho$ being then positive. In this
case, by virtue of sufficiency
of the condition (Theorem 3) we obtain separability
of the state (\ref{st}) for $p={1 \over 2}$.
According  to \cite{renyi} and \cite{tata},
for the above states both the conditions: 2-entropy inequality
\footnote{The $\alpha$-entropy inequalities \cite{renyi} are of the form
$S_\alpha(\varrho)\leq \max_{i=1,2}S_\alpha(\varrho_i)$ for
$1\leq\alpha\leq\infty$. Here $\varrho_i$ are
the reductions of $\varrho$, and $S_\alpha(\varrho)={1\over 1-\alpha}
\ln\text{Tr}
\varrho^\alpha$ for $1<\alpha < \infty$. $S_1$ is the usual
von Neumann entropy and $S_\infty=-\ln\text{Tr}||\varrho||$ \cite{Wehrl}.}
and positivity
of the partial transposition appear to be  equivalent.

Now we will show that the necessary
and sufficient condition of the partial transposition is
essentially stronger.
Consider the states (c.f. \cite{concent}):
\begin{equation}
\varrho=
p|\Psi_{-}\rangle\langle\Psi_{-}| +
(1-p) |\uparrow \uparrow\rangle\langle\uparrow \uparrow|
\label{stn}
\end{equation}
with
$|\Psi_{-}\rangle ={1 \over \sqrt{2}}
(e_1\otimes e_2 - e_2\otimes e_1) \ \  $,
$|\uparrow \uparrow\rangle=e_1\otimes e_1$.
In this case we get
\begin{eqnarray}
\varrho=\left[ \begin{array}{cccc}
          1-p & 0     & 0     & 0   \\
           0    & {p \over 2} & -{p \over 2} & 0       \\
           0    & -{p \over 2} & {p \over 2} & 0       \\
           0    & 0      & 0      & 0
       \end{array}
      \right ], \ \ \
\varrho^{T_2}=\left[ \begin{array}{cccc}
          1-p& 0     & 0     & -{p \over 2}   \\
           0    & {p \over 2}     & 0 & 0     \\
           0    & 0  & { p \over 2} & 0       \\
           -{p \over 2}    & 0      & 0      & 0
       \end{array}
      \right ].
\end{eqnarray}
Here the first of the determinants (\ref{wyzn}) is always positive
but for the second one we have
$W_1=-{p \over 4}^2<0$ for any $p\neq0$.
Hence the state (\ref{stn}) is separable only
if it is a  pure product state $|\uparrow \uparrow\rangle
\langle\uparrow \uparrow\vert$.
On the other hand it can be seen that 2-entropy
inequality is fulfilled for $p\leq{1 \over 3}$.
Thus we see that the necessary and sufficient condition is  essentially
stronger than the 2-entropy one.

In conclusion, we have provided  necessary and sufficient
conditions for separability of
mixed states. In particular, for $2\times 2$ and $2 \times 3$
systems we have obtained an effective criterion of separability.
It completes, in particular, Peres'
result \cite{Peres}.  It follows however, that the positivity
of the partial transposition of the state is not a sufficient
condition in general. Finally, we belive that the above results
will be useful for practical applications of quantum inseparability.

{\it Appendix.- } To prove that $\varrho^{T_2}\geq0$ is not a sufficient condition for
separability in higher dimensions it sufficies to show that the cone $S_T=
\{A\in{\cal A}_1\otimes{\cal A}_2:\ A\geq0,A^{T_2}\geq0\}$ is not equal to the cone
$S=\{\lambda\varrho:\lambda\geq0;\varrho - \text{separable state}\}$
or equivalently to show that the dual cones (the cones $W$ and $W_{T_2}$ of
functionals which are positive on elements from $S$ and $S_T$ respectively)
are not equal. We know that $W$ is isomporphic to the cone of all positive
maps $\Lambda:{\cal A}_1\rightarrow {\cal A}_2$. On the other hand it
can be  seen that
\begin{equation}
W_{T_2}=\{\text{Tr}[A \ \ \cdot \ \ ]: A=B+C^{T_2},\ B,\ C - \text{positive operators}\}.
\end{equation}
Hence $W_{T_2}$ is isomorphic to the cone of all the positive maps of the form
(\ref{cp}). However every positive map $\Lambda:{\cal A}_1\rightarrow {\cal A}_2$
is of the form (\ref{cp}) only for the cases ${\cal A}_1={\cal A}_2={\cal M}_2$,
${\cal A}_1={\cal M}_2$, ${\cal A}_2={\cal M}_3$  or
${\cal A}_1={\cal M}_3$, ${\cal A}_2={\cal M}_2$ \cite{Woronowicz}.
Thus, in general, $W\not=W_{T_2}$ hence the condition $\varrho^{T_2}\geq0$ is not
sufficient for separability of $\varrho$ in higher dimensions.

It is a pleasure for us to thank Asher Peres for useful correspondence
concerning the inseparability problem.

\end{document}